\documentclass[aps,preprint,secnumarabic]{revtex4}
\usepackage{amsmath}
\usepackage{amssymb}
\usepackage{graphicx}
\usepackage{esint}

\begin{document}

\title{Circadian clocks optimally adapt to sunlight for reliable synchronization}

\author{Yoshihiko Hasegawa}

\email[Corresponding author~: ]{hasegawa@cb.k.u-tokyo.ac.jp}

\affiliation{Department of Biophysics and Biochemistry, Graduate School of Science,
The University of Tokyo, Tokyo 113-0033, Japan}

\author{Masanori Arita}

\email[Current address~: Center for Information Biology, National Institute of Genetics, Shizuoka 411-8540, Japan]{}

\affiliation{Department of Biophysics and Biochemistry, Graduate School of Science,
The University of Tokyo, Tokyo 113-0033, Japan}

\affiliation{RIKEN Center for Sustainable Resource Science, Kanagawa 230-0045,
Japan}
\begin{abstract}
Circadian oscillation provides selection advantages through synchronization
to the daylight cycle. However, a reliable clock must be designed
through two conflicting properties: entrainability to synchronize
internal time with periodic stimuli such as sunlight, and regularity
to oscillate with a precise period. These two aspects do not easily
coexist because better entrainability favors higher sensitivity, which
may sacrifice the regularity. To investigate conditions for satisfying
the two properties, we analytically calculated the optimal phase-response
curve with a variational method. Our result indicates an existence
of a dead zone, i.e., a time period during which input stimuli neither
advance nor delay the clock. A dead zone appears only when input stimuli
obey the time course of actual solar radiation but a simple sine curve
cannot yield a dead zone. Our calculation demonstrates that every
circadian clock with a dead zone is optimally adapted to the daylight
cycle. 
\end{abstract}
\maketitle

\section{Introduction\label{sec:introduction}}

Circadian oscillators are prevalent in organisms from bacteria to
human and serve to synchronize bodies with the environmental 24 h
cycle~\cite{Refinetti:2005:CircBook,Gonze:2011:CircadianModel}.
Although the molecular implementation of oscillation is species specific~\cite{Johnsson:2007:CircLightReview,Hubbard:2009:SystCirc,Ukai:2010:MamCircadian,Johnson:2011:CyanoCircadian},
every circadian clocks satisfies two requirements to achieve the reliable
synchronization to the environment: \textbf{entrainability} to synchronize
internal time with periodic stimuli and \textbf{regularity} to oscillate
with a precise period. Circadian clocks are acquired through evolution
independently in bacteria, fungi, plants and animals \cite{Young:2001:GeneticsCirc}.
Nonetheless, entrainability and regularity constitute major characteristics
conserved in all circadian clocks \cite{Johnson:2011:CyanoCircadian},
which strongly suggests that these two properties are essential for
survival. A main source of interference with regularity is discreteness
of molecular species, i.e., molecular noise \cite{Koern:2005:GeneNoiseReview,Perkins:2009:CellDecision,Raser:2010:GeneNoise,Eldar:2010:NoiseRoleGene,Hasegawa:2012:Motor,Viney:2013:AdapNoise}.
Many studies have analyzed the resistance mechanisms of circadian
oscillators against the noise~\cite{Gonze:2002:RobustCircadian,Forger:2005:StocMamCirc,Morelli:2007:CircadianPrecision,Rougemont:2007:PeriCircNoise}.
Regarding entrainability, circadian clocks synchronize their internal
time with the environmental cycle via sunlight, and its effect depends
on the wavelength or fluence, as well as on the phase of the stimulation.
However, entrainability and regularity are conflicting factors, because
circadian clocks with better entrainability are sensitive not only
to the periodic light stimuli, but also to the molecular noise which
interferes with regularity. 

\begin{figure}
\begin{centering}
\includegraphics[width=7cm]{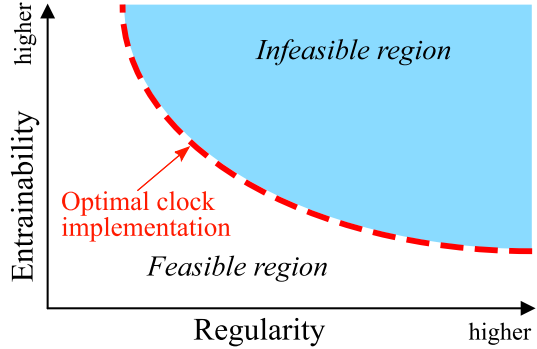}
\par\end{centering}

\caption{Illustrative relation between two tradeoff properties, entrainability
and regularity. There is an infeasible region with respect to entrainability
and regularity (colored area), inside which no clocks can be implemented.
Actual circadian clocks are considered to optimally satisfy them and
such optimal clocks lie on the edge between feasible and infeasible
regions (thick dashed line). \label{fig:pareto_optimal}}
\end{figure}

\begin{figure}
\begin{centering}
\includegraphics[width=13cm]{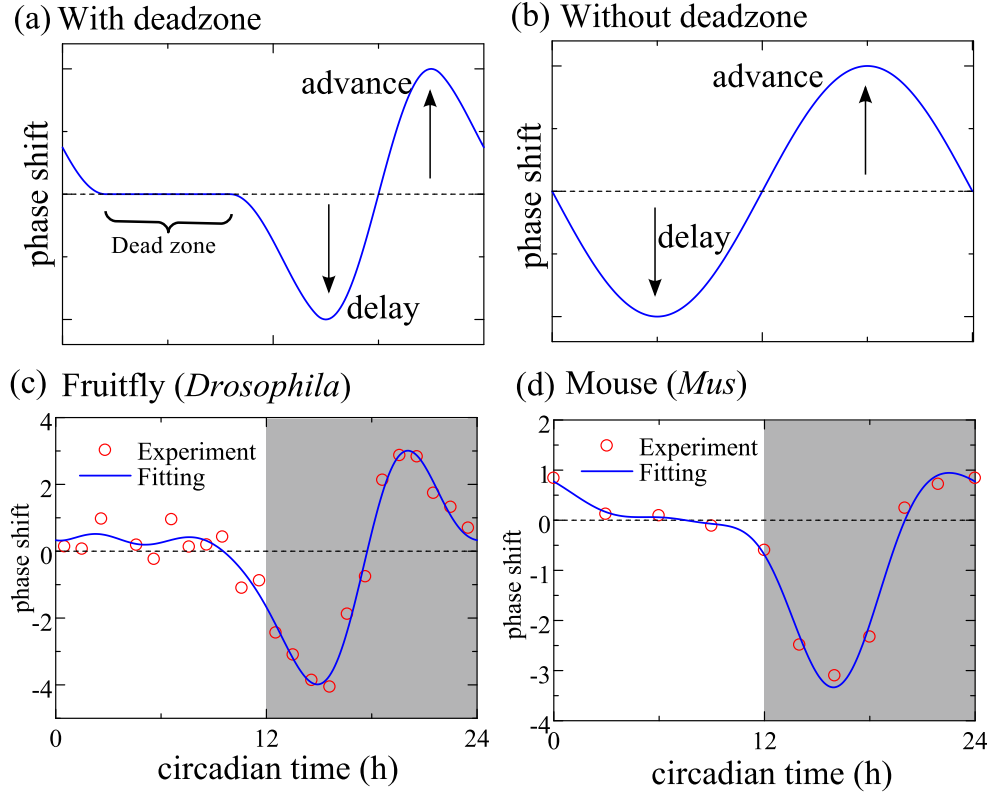} 
\par\end{centering}

\caption{(a)--(b) Illustrations of typical PRCs (a) with and (b) without a
dead zone. (c)--(d) Experimentally observed PRCs as a function of
time (hour) in (c) fruitfly (\emph{Drosophila}) \cite{Hall:1987:BiolRhythm}
and (d) mouse (\emph{Mus}) \cite{Daan:1976:MouseCirc2} with light
pulses (circles) and their trigonometric fitting curves (solid line).
Shaded and nonshaded regions indicate subjective night and day, respectively.
\label{fig:type_1_2_PRCs}}
\end{figure}

Since both regularity and entrainability are important adaptive values,
we expect actual circadian oscillators to optimally satisfy these
two factors (Fig.~\ref{fig:pareto_optimal}). Here we investigate
the optimal phase-response curve (PRC), which is both entrainable
and regular, in the phase oscillator model \cite{Kuramoto:2003:OscBook}
by using the Euler--Lagrange variational method. Our main finding
is the inherent existence of a dead zone in the PRC: optimality is
achieved only when the PRCs have a time period during which light
stimuli neither advance nor delay the clock (Fig.~\ref{fig:type_1_2_PRCs}(a)).
In other words, a PRC with a dead zone (Fig.~\ref{fig:type_1_2_PRCs}(a))
is better adapted than those without a dead zone (Fig.~\ref{fig:type_1_2_PRCs}(b)).
This result is intriguing because a dead zone, with which oscillators
tend to be unaffected by stimuli (i.e. lower entrainability), achieves
better entrainability. We also tested this with two types of input
stimuli: a solar radiation-type input that simulated the time course
of solar radiation intensity (cf. Eq.~\eqref{eq:insolation_input}
and Fig.~\ref{fig:radiation_fig}(a)) and a simple sinusoidal input
(sine curve). Surprisingly, the dead zone in the optimal PRC only
emerges for the solar radiation-type input, not for the sinusoidal
input. Many experimental studies reported the existence of a dead
zone in various species (Figs.~\ref{fig:type_1_2_PRCs}(c) and (d)
show experimentally observed PRCs of (c) fruitfly \cite{Hall:1987:BiolRhythm}
and (d) mouse \cite{Daan:1976:MouseCirc2}, respectively). Our results
indicate that circadian oscillators in various species have adapted
to solar radiation for reliable synchronization. 

\section{Models and methods}

\begin{figure}
\begin{centering}
\includegraphics[width=13cm]{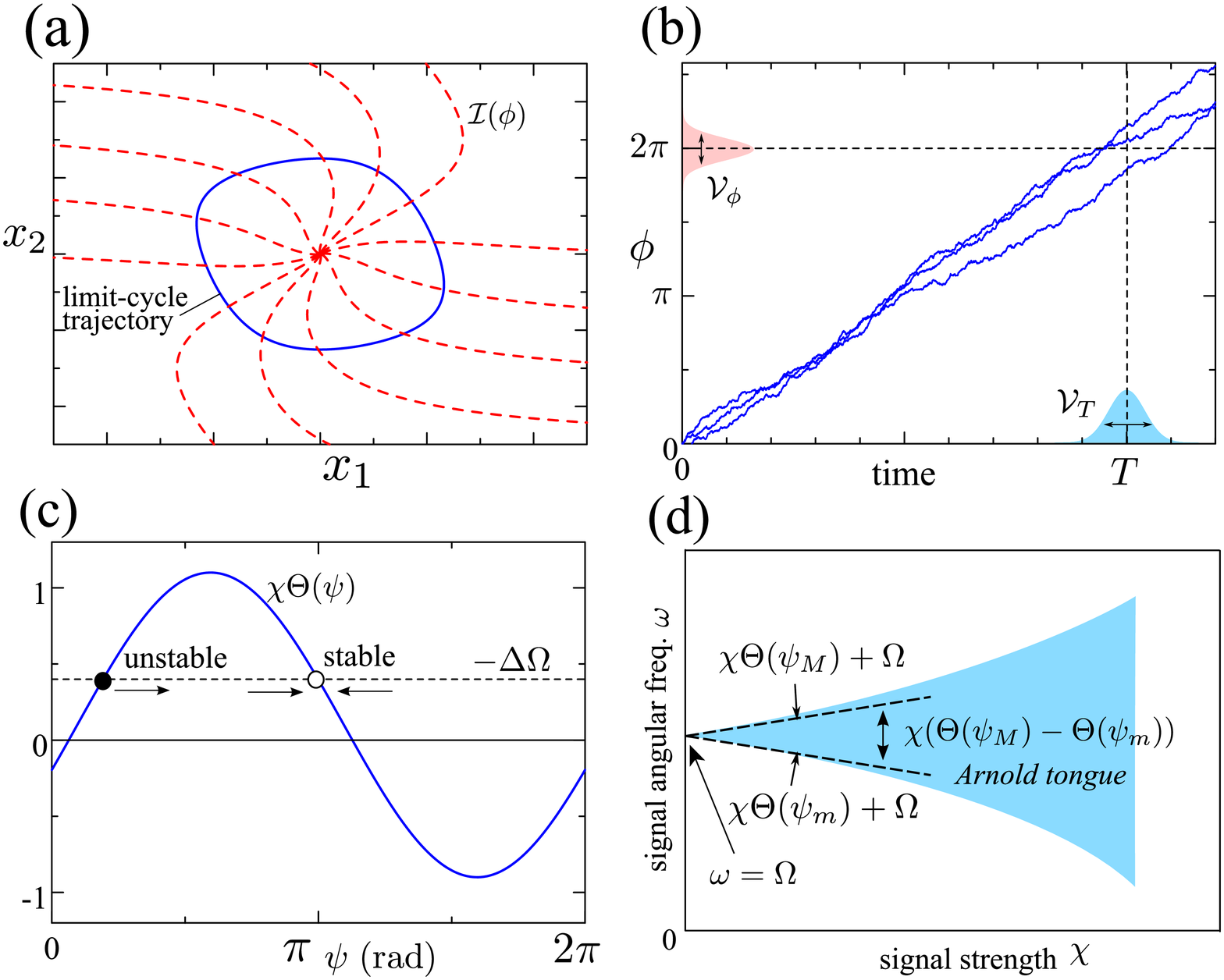}
\par\end{centering}

\caption{(a) Illustration of the isochron $\mathcal{I}(\phi)$, the solid and
dashed lines describing a limit-cycle trajectory and its isochron
drawn at intervals of $\pi/6$, respectively. (b) Relation between
the phase variance $\mathcal{V}_{\phi}$ and the period variance $\mathcal{V}_{T}$
in Langevin equation~\eqref{eq:LC_def} (the solid lines represent
trajectories of the Langevin equation). $\mathcal{V}_{\phi}$ is the
variance of the phase $\phi$ at time $t=T$ and $\mathcal{V}_{T}$
is the variance of the first passage time from $0$ to $2\pi$, which
can be approximated by $\mathcal{V}_{T}\simeq\mathcal{V}_{\phi}T^{2}/(2\pi)^{2}$.
(d) Arnold tongue (colored region), which shows the parameter region
for synchronization to an input signal, with respect to the signal
angular frequency $\omega$ (vertical axis) and the signal strength
$\chi$ (horizontal axis). The dashed line is a linear approximation
(Eq.~\eqref{eq:phase_lock_condition}) of the border of the Arnold
tongue when the input strength $\chi$ is sufficiently small. \label{fig:intuitive_fig}}
\end{figure}

\subsection{Phase oscillator model}

Circadian oscillators basically comprise interaction between mRNAs
and proteins, whose dynamics can be modeled by differential equations.
A circadian oscillator of $N$-molecular species can be represented
by 
\begin{equation}
\frac{dx_{i}}{dt}=F_{i}(\boldsymbol{x})\hspace{1em}(i=1,2,\cdots,N),\label{eq:mass_action}
\end{equation}
where the $N$-dimensional vector $\boldsymbol{x}=(x_{1},x_{2},\cdots,x_{N})$
denotes the concentration of molecular species (mRNAs or proteins).
The effect of noise on genetic oscillators has been a subject of considerable
interest, and noise-resistant mechanisms have been extensively studied
\cite{Gonze:2002:RobustCircadian,Forger:2005:StocMamCirc,Vilar:2002:NoiseResistGeneOsc,Scott:2006:ExtrinsicNoise,Yoda:2007:NoiseOsc,Morelli:2007:CircadianPrecision,Rougemont:2007:PeriCircNoise}.
In general, the dynamics of the $i$-th molecular concentration in
a circadian oscillator subject to molecular noise is described by
the following Langevin equation (Stratonovich interpretation): 
\begin{equation}
\frac{dx_{i}}{dt}=F_{i}(\boldsymbol{x};\rho)+Q_{i}(\boldsymbol{x})\xi_{i}(t),\label{eq:LC_def}
\end{equation}
where $Q_{i}(\boldsymbol{x})$ is an arbitrary function representing
the multiplicative terms of the noise, $\xi_{i}(t)$ is white Gaussian
noise with the correlation $\left\langle \xi_{i}(t)\xi_{j}(t^{\prime})\right\rangle =2\delta_{ij}\delta(t-t^{\prime})$
(a bracket $\left\langle \cdot\right\rangle $ denotes expectation),
and $\rho$ is a model parameter.

Circadian oscillators synchronize to environmental cycles by responding
to a periodic input signal (light stimuli). We let $\rho$ in Eq.~\eqref{eq:LC_def}
be stimulated by the input signal: for example, $\rho$ can be the
degradation rate (For simplicity, we consider that the input signal
affects only one parameter). We use Eq.~\eqref{eq:LC_def} for calculating
regularity and entrainability of circadian oscillators.

\subsection{Definition of regularity}

Because the circadian oscillator of Eq.~\eqref{eq:LC_def} is subject
to noise, its period varies cycle to cycle. We use the term regularity
for the period variance of the oscillation (higher regularity corresponds
to smaller period variance). Let us first consider the case without
input signals (i.e., $\rho$ is constant).  As Eq.~\eqref{eq:mass_action}
exhibits periodic oscillation, we can naturally define the phase $\phi\in[0,2\pi)$
on Eq.~\eqref{eq:mass_action} by
\begin{equation}
\frac{d\phi}{dt}=\Omega,\label{eq:OmegaODE}
\end{equation}
where $\Omega=2\pi/T$ is the angular frequency of the oscillation
($T$ is a period of the oscillation). The phase $\phi$ in Eq.~\eqref{eq:OmegaODE}
is only defined on a closed orbit of the unperturbed limit-cycle oscillation.
However, we can expand the definition into the entire $\boldsymbol{x}=(x_{1},x_{2},\cdots,x_{N})$
space, where the equiphase surface is referred to as the isochron
$\mathcal{I}(\phi)$ (Fig.~\ref{fig:intuitive_fig}(a)). By using
standard stochastic phase reduction \cite{Kuramoto:2003:OscBook},
Eq.~\eqref{eq:LC_def} can be transformed into the following Langevin
equation with respect to the phase variable $\phi$ (Stratonovich
interpretation): 
\begin{equation}
\frac{d\phi}{dt}=\Omega+\sum_{i=1}^{N}U_{i}(\phi)Q_{i}(\phi)\xi_{i}(t),\label{eq:phi_Langevin}
\end{equation}
where $\boldsymbol{U}(\phi)=(U_{1}(\phi),\cdots,U_{N}(\phi))$ is
an infinitesimal PRC (iPRC) $\boldsymbol{U}(\phi)=\left.\nabla_{\boldsymbol{x}}\phi\right|_{\boldsymbol{x}=\boldsymbol{x}_{\mathrm{LC}}(\phi)}$,
and we abbreviated $Q_{i}(\boldsymbol{x}_{\mathrm{LC}}(\phi))$ as
$Q_{i}(\phi)$. iPRC $U_{i}(\phi)$ quantifies the extent of phase
advance or delay when perturbed along an $x_{i}$ coordinate direction
at phase $\phi$. The $N$-dimensional vector $\boldsymbol{x}_{\mathrm{LC}}(\phi)$
denotes a point on the limit-cycle trajectory at phase $\phi$, where
LC stands for limit cycle. The value of iPRC $U_{i}(\phi)$ is calculated
as a solution of an adjoint equation~\cite{Izhikevich:2007:NeuroBook}
or as the set of eigenvectors of a monodromy matrix in the Floquet
theory \cite{Kuramoto:2003:OscBook} for arbitrary oscillators. Let
$P(\phi;t)$ be the probability density function of $\phi$ at time
$t$. From Eq.~\eqref{eq:phi_Langevin}, the Fokker--Planck equation
(FPE) \cite{Risken:1989:FPEBook} of $P(\phi;t)$ is given by 
\begin{equation}
\frac{\partial P(\phi;t)}{\partial t}=\left\{ -\frac{\partial}{\partial\phi}\left(\Omega+\mathcal{F}(\phi)\right)+\frac{\partial^{2}}{\partial\phi^{2}}\mathcal{G}(\phi)\right\} P(\phi;t),\label{eq:FPE}
\end{equation}
where 
\begin{eqnarray}
\mathcal{F}(\phi) & = & \sum_{i=1}^{N}U_{i}(\phi)Q_{i}(\phi)\frac{d}{d\phi}U_{i}(\phi)Q_{i}(\phi),\label{eq:FPE_F_def}\\
\mathcal{G}(\phi) & = & \sum_{i=1}^{N}U_{i}(\phi)^{2}Q_{i}(\phi)^{2}.\label{eq:FPE_G_def}
\end{eqnarray}
Introducing a slow variable $\varphi=\phi-\Omega t$, the FPE of the
probability density function $\Pi(\varphi;t)=P(\phi=\varphi+\Omega t;t)$
is given by 
\begin{equation}
\frac{\partial}{\partial t}\Pi(\varphi;t)=\left\{ -\frac{\partial}{\partial\varphi}\mathcal{F}(\varphi+\Omega t)+\frac{\partial^{2}}{\partial\varphi^{2}}\mathcal{G}(\varphi+\Omega t)\right\} \Pi(\varphi;t).\label{eq:Pi_FPE}
\end{equation}
With sufficiently weak noise, $\Pi(\varphi;t)$ is a slowly fluctuating
function of $t$. In such cases, $\mathcal{F}(\varphi+\Omega t)$
and $\mathcal{G}(\varphi+\Omega t)$ fluctuate much faster than $\Pi(\varphi;t)$,
thus these two terms can be averaged for one period while keeping
$\Pi(\varphi;t)$ constant (phase averaging). In other words, we separate
time scales between $\mathcal{F}(\varphi+\Omega t)$, $\mathcal{G}(\varphi+\Omega t)$
and $\Pi(\varphi;t)$. By phase averaging, $\mathcal{F}(\varphi+\Omega t)$
vanishes because of the periodicity (use integration by parts), yielding
\begin{equation}
\frac{\partial}{\partial t}\Pi(\varphi;t)=D\frac{\partial^{2}}{\partial\varphi^{2}}\Pi(\varphi;t),\label{eq:diffusion_equation}
\end{equation}
with 
\begin{equation}
D=\frac{1}{2\pi}\int_{0}^{2\pi}d\theta\,\sum_{i=1}^{N}U_{i}(\theta)^{2}Q_{i}(\theta)^{2}.\label{eq:def_D}
\end{equation}
Please see Ref.~\cite{Kuramoto:2003:OscBook} for further details
of stochastic phase reduction and the phase-averaging procedure. From
Eq.~\eqref{eq:diffusion_equation}, because $\Pi(\varphi=\phi-\Omega t;t)[=P(\phi;t)]$
obeys a simple one-dimensional diffusion equation, its solution is
represented by 
\begin{equation}
P(\phi;t)=\frac{1}{\sqrt{4\pi Dt}}\exp\left(-\frac{(\phi-\Omega t)^{2}}{4Dt}\right).\label{eq:P_phi_t}
\end{equation}
Equation~\eqref{eq:P_phi_t} shows that the variance of the phase
after one period $T$ is 
\[
\mathcal{V}_{\phi}=2DT.
\]
In Eq.~\eqref{eq:phi_Langevin}, the average period corresponds to
the mean first passage time with $\phi$ starting from $0$ to $2\pi$,
and the period variance is the variance of the first passage time.
Because direct calculation of the period variance is difficult, we
approximate the period variance $\mathcal{V}_{T}$ with the phase
variance $\mathcal{V}_{\phi}$, after Kori \emph{et. al} \cite{Kori:2012:OscReg}.
As the phase $\phi$ crosses a threshold $\phi=2\pi$ with gradient
$2\pi/T$ without noise, there is a scaling relation $\sqrt{\mathcal{V}_{T}}\simeq\sqrt{\mathcal{V}_{\phi}}T/(2\pi)$
for sufficiently weak noise \cite{Kori:2012:OscReg} (Fig.~\ref{fig:intuitive_fig}(b)).
Consequently, the variance of the period is approximated by 
\begin{equation}
\mathcal{V}_{T}\simeq\mathcal{V}_{\phi}\left(\frac{T}{2\pi}\right)^{2}=\frac{T^{3}}{4\pi^{3}}\int_{0}^{2\pi}d\theta\,\sum_{i=1}^{N}U_{i}(\theta)^{2}Q_{i}(\theta)^{2}.\label{eq:var_T}
\end{equation}

\subsection{Definition of entrainability}

The entrainment property is an important characteristic of limit-cycle
oscillators and attracts attention in systems biology \cite{Gonze:2000:Entrainment,Hasty:2002:GeneOsc,Begheri:2008:RobustEntrain,Abraham:2010:CoupleEntrain,Hasegawa:2012:GeneOsc,Erzberger:2013:CircEntrain}.
For instance, a period mismatch in coupled oscillators is known to
enhance entrainability in genetic oscillators \cite{Hasegawa:2012:GeneOsc}.
Light stimuli affect the rate constants, i.e., the parameter $\rho$
in Eq.~\eqref{eq:LC_def} is perturbed as $\rho+d\rho$ by the input
signal. Equation~\eqref{eq:LC_def} can be viewed as representing
the dynamics of a tilted periodic potential (i.e., ratchet) subject
to noise. Since a synchronizable condition corresponds to the existence
of stable points in the ratchet-like potential, the entrainability
can be discussed without considering the noise. Consequently, in contrast
to the calculation of regularity, in the evaluation of the entrainability,
we consider a case without molecular noise (i.e., $Q_{i}(\boldsymbol{x})=0$
in Eq.~\eqref{eq:LC_def}).

Let $p(\omega t)$ be an input signal with angular frequency $\omega$.
Considering a weak periodic input signal $d\rho=\chi p(\omega t)$,
where $\chi$ is the signal strength ($\chi\ge0$), and applying the
phase reduction approach to Eq.~\eqref{eq:LC_def}, the time evolution
of the phase variable $\phi$ is given by
\begin{eqnarray}
\frac{d\phi}{dt} & = & \Omega+\sum_{i=1}^{N}\frac{\partial\phi}{\partial x_{i}}\frac{\partial F_{i}(\phi;\rho)}{\partial\rho}d\rho,\nonumber \\
 & = & \Omega+\chi Z(\phi)p(\omega t),\label{eq:phase_reduction_def}
\end{eqnarray}
with $F_{i}(\phi;\rho)=F_{i}(\boldsymbol{x}_{\mathrm{LC}}(\phi);\rho)$
and 
\begin{equation}
Z(\phi)=\sum_{i=1}^{N}U_{i}(\phi)\frac{\partial F_{i}(\phi;\rho)}{\partial\rho},\label{eq:PRC_Z}
\end{equation}
where $Z(\phi)$ is the PRC with respect to the parameter $\rho$
and corresponds to experimentally observed PRCs. In order to distinguish
$Z(\phi)$ from iPRC $U_{i}(\phi)$, we will refer to $Z(\phi)$ as
the \emph{parametric PRC} (pPRC) \cite{Taylor:2008:Sensitivity}.
Note that the definition of measured PRCs are different from pPRCs
$Z(\phi)$ in a rigorous definition; the experimentally measured PRCs
quantify the phase shift $\Delta\phi$ caused by light stimuli while
pPRCs $Z(\phi)$ are normalized by the strength of perturbation, i.e.
$Z(\phi)=\partial\phi/\partial\rho\simeq\Delta\phi/\Delta\rho$. Therefore,
the range of the measured pPRCs have limitation $-\pi\le\Delta\phi<\pi$
while pPRCs $Z(\phi)$ do not. The phase reduction can yield reliable
results only when the perturbed trajectory is close to the unperturbed
limit-cycle trajectory (i.e., $\chi$ is sufficiently small). 

We next evaluate the extent of synchronization to the periodic input
signal. By introducing another slow variable $\psi=\phi-\omega t$
in Eq.~\eqref{eq:phase_reduction_def}, we can again apply the phase-averaging
procedure, which yields 
\begin{equation}
\frac{d\psi}{dt}=\Delta\Omega+\chi\Theta(\psi),\label{eq:slow_dyn}
\end{equation}
with $\Delta\Omega=\Omega-\omega$ and 
\begin{equation}
\Theta(\psi)=\frac{1}{2\pi}\int_{0}^{2\pi}d\theta\, Z(\psi+\theta)p(\theta).\label{eq:Theta_def}
\end{equation}
The oscillator of interest can synchronize to input signals when there
is a stable solution of $\psi$ in $\dot{\psi}=0$ (Eq.~\eqref{eq:slow_dyn}).
The stable solution is an intersection point of $\Theta(\psi)$ and
$-\Delta\Omega$ with $d\Theta/d\psi<0$ (an empty circle in Fig.~\ref{fig:intuitive_fig}(c)).
Then a condition for the existence of a stable solution is 
\begin{equation}
\chi\Theta(\psi_{m})+\Omega<\omega<\chi\Theta(\psi_{M})+\Omega,\label{eq:phase_lock_condition}
\end{equation}
where $\psi_{M}=\mathrm{argmax}_{\psi}\Theta(\psi)$ and $\psi_{m}=\mathrm{argmin}_{\psi}\Theta(\psi)$. 

We define entrainability, the extent of synchronization to the periodic
input signal, by the width of the Arnold tongue, which is a domain
with respect to $\chi$ (signal strength) and $\omega$ (signal angular
frequency). The shaded region in Fig.~\ref{fig:intuitive_fig}(d)
represents the Arnold tongue; with parameters $\chi$ and $\omega$
inside the Arnold tongue, the oscillator can synchronize to a periodic
input signal. Because Eq.~\eqref{eq:phase_lock_condition} constitutes
a linear approximation of the Arnold tongue for sufficiently small
$\chi$, the width of the Arnold tongue is given by $\chi\left(\Theta(\psi_{M})-\Theta(\psi_{m})\right)$
under the linear approximation. Thus we define the entrainability
$\mathcal{E}$, or the extent of synchronization, as 
\begin{equation}
\mathcal{E}=\Theta(\psi_{M})-\Theta(\psi_{m}).\label{eq:entrainabilit_def}
\end{equation}
Intuitively, a circadian oscillator with better entrainability (i.e.,
larger $\mathcal{E}$) can synchronize to an input signal that has
a period further from that of the oscillator. The calculation above
is standard in the phase reduction approach, and further details are
available in Ref.~\cite{Kuramoto:2003:OscBook}.

\subsection{Variational method}

We use the variational method to calculate the optimal PRCs, which
maximize the entrainability $\mathcal{E}$ subject to constant variance
$\mathcal{V}_{T}=\sigma_{T}^{2}$ (the optimal solutions correspond
to the edge in Fig.~\ref{fig:pareto_optimal}, which is described
by the thick dashed line). The constrained optimization of $U_{i}(\phi)$
can be intuitively interpreted as maximization of weighted area (Eq.~\eqref{eq:entrainabilit_def}),
where the input being the weight, with constant area under the squared
magnitude (Eq.~\eqref{eq:var_T}). In a simple term, the optimality
is reached when the magnitude of the PRC is small during intervals
when the input magnitude is small (and vice versa). In the context
of neuronal oscillators, a study \cite{Abouzeid:2009:OptimalPRC}
has used the variational method to calculate the optimal PRCs for
stochastic synchrony (noise-induced synchronization \cite{Teramae:2004:NoiseIndSync,Nakao:2007:NISinLC}).

The variational equation to be optimized is 
\begin{equation}
\mathcal{L}[\boldsymbol{U}]=\mathcal{E}[\boldsymbol{U}]-\lambda\mathcal{V}_{T}[\boldsymbol{U}],\label{eq:variational_L2}
\end{equation}
where $\lambda$ is the Lagrange multiplier. Note that variational
equation~\eqref{eq:variational_L2} is similar to Ref.~\cite{Harada:2010:OptimalInput},
which optimizes the input signal for the maximal entrainment under
constant power of the input. The variational condition $\delta\mathcal{L}[\boldsymbol{U}]=0$
yields the optimal iPRC 
\begin{equation}
U_{i}(\phi)=\frac{\pi^{2}}{T^{3}\lambda}\frac{p(\phi-\psi_{M})-p(\phi-\psi_{m})}{Q_{i}(\phi)^{2}}\frac{\partial F_{i}(\phi;\rho)}{\partial\rho},\label{eq:iPRC_solution}
\end{equation}
and the pPRC is calculated with Eq.~\eqref{eq:PRC_Z}:
\begin{equation}
Z(\phi)=\frac{\pi^{2}}{T^{3}\lambda}\sum_{i=1}^{N}\frac{p(\phi-\psi_{M})-p(\phi-\psi_{m})}{Q_{i}(\phi)^{2}}\left(\frac{\partial F_{i}(\phi;\rho)}{\partial\rho}\right)^{2}.\label{eq:pPRC_solution}
\end{equation}
Since $\psi_{M}$ and $\psi_{m}$ themselves depend on $U_{i}(\phi)$,
they have to satisfy a self-consistent condition, i.e., Eq.~\eqref{eq:entrainabilit_def}
is maximal with $\psi_{M}$ and $\psi_{m}$. Consequently, we maximize
the following function: 
\begin{equation}
\mathcal{E}(\Delta,\delta)=\sqrt{\frac{\pi\sigma_{T}^{2}}{T^{3}}\Psi(\Delta,\delta)},\label{eq:F_final}
\end{equation}
with 
\begin{equation}
\Psi(\Delta,\delta)=\int_{0}^{2\pi}d\theta\sum_{i=1}^{N}\frac{\left(p(\theta-\Delta)-p(\theta)\right)^{2}}{Q_{i}(\theta+\delta)^{2}}\left(\frac{\partial F_{i}(\theta+\delta;\rho)}{\partial\rho}\right)^{2},\label{eq:Psi_def}
\end{equation}
where $\Delta=\psi_{M}-\psi_{m}$ and $\delta=\psi_{m}$. The optimal
iPRC can be obtained by first finding the maximum solution of $\Psi(\Delta,\delta)$
with respect to $\Delta$ and $\delta$, and then substituting the
obtained solution $\psi_{m}=\delta$ and $\psi_{M}=\delta+\Delta$
into Eqs.~\eqref{eq:iPRC_solution} and \eqref{eq:pPRC_solution}.

\subsection{Input signal of solar radiation model}

\begin{figure}
\begin{centering}
\includegraphics[width=13cm]{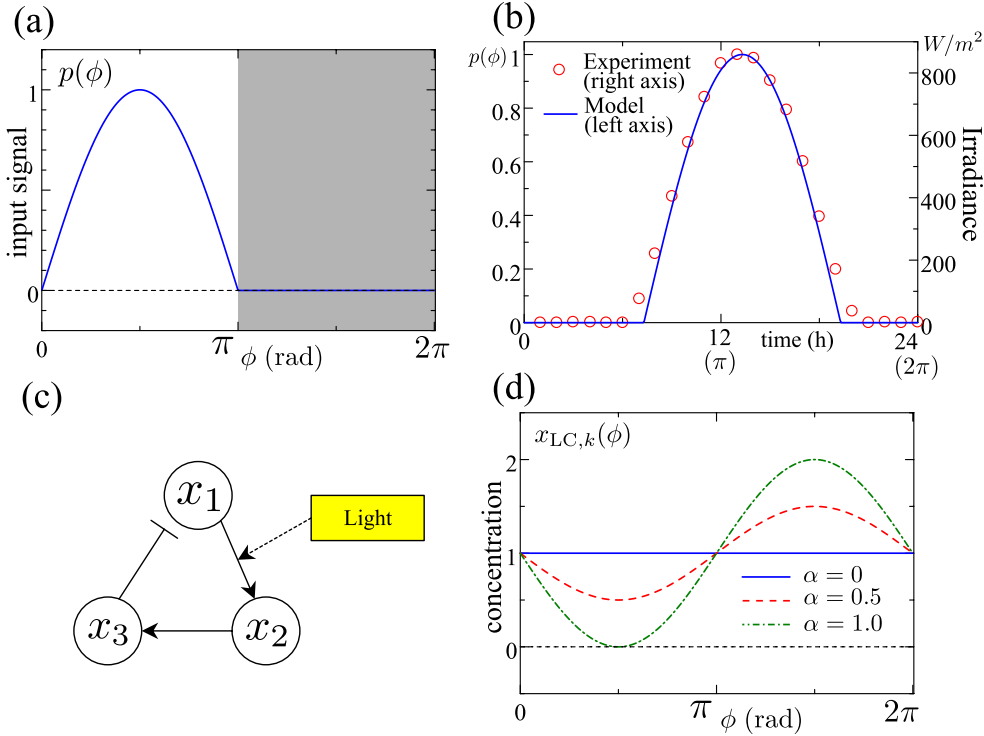} 
\par\end{centering}

\caption{(a) Solar radiation input of Eq.~\eqref{eq:insolation_input}, where
the shaded region denotes night. (b) Comparison between solar radiation
input (Eq.~\eqref{eq:insolation_input}) and actual observed irradiance
data taken from Ref.~\cite{Vick:2013:SolPowPlants}. The solar radiation
input (solid line) and the observed data (circle) refer to left and
right axes, respectively. A unit of the observed data is watt per
square meter ($W/m^{2}$). (c) Gene regulatory circuit of hypothetical
circadian clock. In this example, $x_{1}$ and $x_{2}$ describe mRNA
and protein, respectively, and $x_{3}$ represses the transcription
of $x_{1}$. Light stimuli increases the translational efficiency.
(d) Time course of $x_{\mathrm{LC},k}(\phi)$ (Eq.~\eqref{eq:trajectory_assumption}),
which is a variable to be multiplied by the parameter $\rho$ (Eq.~\eqref{eq:linear_effects}).
\label{fig:radiation_fig}}
\end{figure}

Optimal PRCs depend on input signals, as seen in Eqs.~\eqref{eq:iPRC_solution}
and \eqref{eq:pPRC_solution}. The most common synchronizer in circadian
oscillators is sunlight, for which the strength is determined by 24
h-periodic solar irradiance. The solar irradiance is calculated by
$I=I_{0}\cos\vartheta$ and $I=0$ when the sun is above the horizon
($0\le\vartheta<\pi$) and below the horizon ($\pi\le\vartheta<2\pi$),
respectively, where $\vartheta$ is the zenith angle and $I_{0}$
is the maximum irradiance \cite{Hartmann:1994:PhysClimate}. It can
be approximated by 
\begin{equation}
p(\omega t)=\mathrm{ramp}(\sin(\omega t)),\label{eq:insolation_input}
\end{equation}
where $\mathrm{ramp}(x)$ is the ramp function defined by $\mathrm{ramp}(x)=x$
for $x\ge0$ and $\mathrm{ramp}(x)=0$ for $x<0$. We call Eq.~\eqref{eq:insolation_input}
the \emph{solar radiation input}, whose plot is shown in Fig.~\ref{fig:radiation_fig}(a)
(the shaded region represents night). In order to show the validity
of the solar radiation modeling, we compare Eq.~\eqref{eq:insolation_input}
with observed irradiance data from Ref.~\cite{Vick:2013:SolPowPlants},
which are shown in a dual axis plot of Fig.~\ref{fig:radiation_fig}(b).
In Fig.~\ref{fig:radiation_fig}(b), Eq.~\eqref{eq:insolation_input}
is plotted by the solid line (left axis) and the observed data by
the dashed line (right axis), where a unit of the observed data is
watt per square meter ($W/m^{2}$). The solar radiation input of Eq.~\eqref{eq:insolation_input}
is shifted horizontally so that Eq.~\eqref{eq:insolation_input}
becomes a good fit to the data. From Fig.~\ref{fig:radiation_fig}(b),
the solar radiation input is in good agreement with the observed data,
which verifies the validity of Eq.~\eqref{eq:insolation_input} as
a solar radiation model. 

For comparison, we also employ a sinusoidal input, which is common
in nonlinear sciences: 
\begin{equation}
p(\omega t)=\sin(\omega t).\label{eq:sin_input}
\end{equation}
Note that $p(\omega t)=B+\sin(\omega t)$, where $B$ is an arbitrary
constant, also yields the same optimal PRCs as Eq.~\eqref{eq:sin_input}
because a constant $B$ in the signal is offset in Eqs.~\eqref{eq:iPRC_solution}--\eqref{eq:Psi_def}.
Although a constant $B$ does not play any roles in formation of the
optimal PRCs, different $B$ result in different Arnold tongues in
general. For calculating the optimal PRCs, we use Eqs.~\eqref{eq:insolation_input}
and \eqref{eq:sin_input}.

\section{Results}

\subsection{Optimal PRC of solar radiation input}

\begin{figure*}
\begin{centering}
\includegraphics[width=16cm]{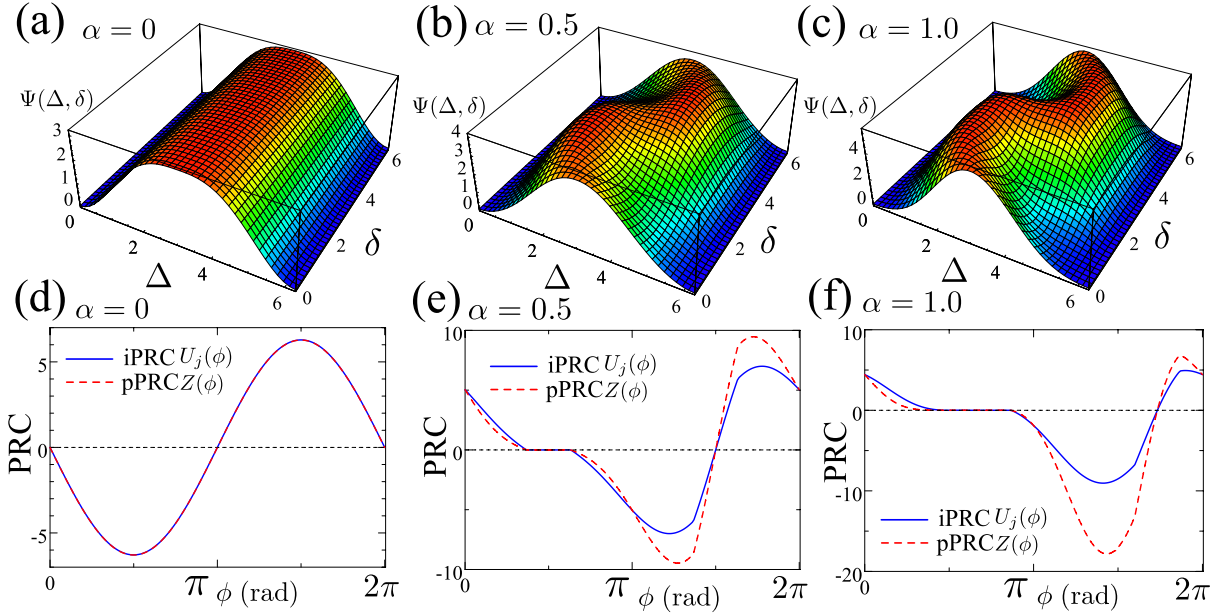} 
\par\end{centering}

\caption{(a)--(c) Landscape of $\Psi(\Delta,\delta)$ as a function of $\Delta$
and $\delta$ with solar radiation input (Eq.~\eqref{eq:insolation_input})
for (a) $\alpha=0$, (b) $\alpha=0.5$, and (c) $\alpha=1.0$, where
the maximum points are parameters for the optimal PRC. (d)--(f) Optimal
PRCs with solar radiation input: (d) $\alpha=0$, (e) $\alpha=0.5$,
and (f) $\alpha=1.0$. In (d)--(f), the solid and dashed lines denote
iPRCs $U_{j}(\phi)$ and pPRCs $Z(\phi)$, respectively (in (d), solid
and dashed lines are indistinguishable). The maximal parameters $(\Delta,\delta)$
for (d)--(f) are (d) $(\pi,0)$, (e) $(2.31,1.99)$, and (f) $(2.30,2.72)$.
In (d), a parallel shift of the PRC is also optimal ($\delta$ can
be an arbitrary value). In (e), symmetric PRCs with respect to the
horizontal axis are also optimal. In (f), symmetric PRCs with respect
to the horizontal axis or $\phi=3\pi/2$ are also optimal (see the
text). The pPRCs correspond to experimentally observed PRCs. \label{fig:optimal_insolation} }
\end{figure*}

Light stimuli generally affect the oscillatory dynamics multiplicatively,
i.e., they act on the rate constants or transcriptional efficiency
of the gene regulatory circuits \cite{Johnsson:2007:CircLightReview,Golombek:2010:CircadianEntrainment}.
We assume that the $j$-th molecular species includes a parameter
$\rho$ as 
\begin{equation}
F_{j}(\boldsymbol{x};\rho)=\tilde{F}_{j}(\boldsymbol{x})+\rho x_{k},\label{eq:linear_effects}
\end{equation}
where $\tilde{F}_{j}(\boldsymbol{x})$ represents the terms that do
not include $\rho$, and $x_{k}$ is the concentration of the $k$-th
molecular species. Here, $k\in\{1,2,\cdots,N\}$ can take any value
regardless of $j$ (both $j\ne k$ and $j=k$ are allowed). For example,
let Fig.~\ref{fig:radiation_fig}(c) be a gene regulatory circuit
of a hypothetical circadian clock, where symbols $\rightarrow$ and
$\dashv$ represent positive and negative regulations and $x_{i}$
are molecular species (please see Ref.~\cite{Novak:2008:BiochemOsc}
for typical motifs of biochemical oscillators). Suppose $x_{1}$ and
$x_{2}$ are mRNA and corresponding protein, respectively, and light
stimuli increase the translational efficiency. In this case, the dynamics
of light entrainment can be described by Eq.~\eqref{eq:linear_effects}
with $j=2$, $k=1$ and $\rho$ being the translation rate. In Eq.~\eqref{eq:linear_effects},
although we can also consider an alternative case $F_{j}(\boldsymbol{x};\rho)=\tilde{F}_{j}(\boldsymbol{x})-\rho x_{k}$
(a negative sign), the optimal pPRCs remain unchanged under the inversion
which is seen from Eqs.~\eqref{eq:pPRC_solution} and \eqref{eq:Psi_def}.
Consequently, we only consider the positive case to calculate the
optimal PRCs (i.e. Eq.~\eqref{eq:linear_effects}). However, note
that relation between iPRCs and pPRCs are affected by the inversion
of the sign, and the difference matters when considering biological
feasibility. 

When using phase reduction, the dynamics of the limit cycle are considered
on the unperturbed limit-cycle trajectories $\boldsymbol{x}_{\mathrm{LC}}$,
and hence the points on the limit cycle can be uniquely determined
by the phase $\phi$. Consequently, under the phase reduction, $x_{k}$
is replaced by $x_{\mathrm{LC},k}(\phi)$ in Eq.~\eqref{eq:linear_effects},
where $x_{\mathrm{LC},k}(\phi)$ is the $k$-th coordinate of $\boldsymbol{x}_{\mathrm{LC}}$
(i.e., $\partial_{\rho}F_{j}(\phi;\rho)=x_{\mathrm{LC},k}(\phi)$
in Eq.~\eqref{eq:iPRC_solution}). Here, $x_{\mathrm{LC},k}(\phi)$
corresponds to the time course of the concentration of the $k$-th
molecular species. Because $x_{\mathrm{LC},k}(\phi)$ constitutes
a core clock component and is generally a smooth $2\pi$-periodic
function, we approximate it with a sinusoidal function: 
\begin{equation}
x_{\mathrm{LC},k}(\phi)=1-\alpha\sin\left(\phi+u\right),\label{eq:trajectory_assumption}
\end{equation}
where $u$ is the initial phase and $\alpha$ denotes the amplitude
of the oscillation (Fig.~\ref{fig:radiation_fig}(d)). To ensure
$x_{\mathrm{LC},k}(\phi)\ge0$, we set $0\le\alpha\le1$, and $\alpha=0$
recovers the additive case. Since the initial phase $u$ does not
play any role ($u$ is offset by $\delta$ in Eq.~\eqref{eq:Psi_def})
when the white Gaussian noise is additive (i.e., $Q_{i}(\boldsymbol{x})\propto1$),
we also set $u=0$. The parametric approximation of Eq.~\eqref{eq:trajectory_assumption}
enables an almost closed form for the overall calculations. Although
we assumed in Eq.~\eqref{eq:linear_effects} that effects of $\rho$
only depend on $x_{k}$, we can generalize Eq.~\eqref{eq:linear_effects}
to $F_{j}(\boldsymbol{x};\rho)=\tilde{F}_{j}(\boldsymbol{x})+\rho K(\boldsymbol{x})$
where $K(\boldsymbol{x})$ is a nonlinear function (a $2\pi$-periodic
function) and is assumed to be well approximated by $1-\alpha\sin(\phi+u)$.
By this generalization, our theory can be applied to other possible
light entrainment mechanisms such as the inter-cellular coupling \cite{Ohta:2005:LightDesync}.
Our model only needs details about molecular species which have light
input entry points but not about a whole molecular network. However,
this advantage in turn means that we can not specify iPRCs $U_{i}(\phi)$
of molecular species not having light input entry points. Consequently,
for a noise term $Q_{i}(\boldsymbol{x})$, we assume that the white
Gaussian noise is additive and is present only in the $j$-th coordinate
equation ($Q_{j}(\phi)=\sqrt{q}$, where $q$ is the noise intensity
and $Q_{i}(\phi)=0$ for $i\ne j$).

Figures~\ref{fig:optimal_insolation}(a)--(c) show the landscape
of $\Psi(\Delta,\delta)$ as functions of $\Delta$ and $\delta$,
and Figs.~\ref{fig:optimal_insolation}(d)--(f) express the optimal
iPRCs $U_{j}(\phi)$ and pPRCs $Z(\phi)$ for the solar radiation
input (an explicit expression of $\Psi(\Delta,\delta)$ is given in
Appendix A). The optimal PRC shape does not depend on the model parameters
such as the period $T$, its variance $\sigma_{T}^{2}$, or noise
intensity $q$. These three parameters only act on the magnitude of
the PRCs (i.e., the vertical scaling of the PRCs). Consequently, we
normalized $T=1$, $\sigma_{T}^{2}=1$, and $q=1$, as shown in Fig.~\ref{fig:optimal_insolation}.
As the optimal PRCs depend on $\alpha$, $\Psi(\Delta,\delta)$ is
plotted for three cases: (a) $\alpha=0$, (b) $\alpha=0.5$, and (c)
$\alpha=1.0$, where the maximal points $(\Delta,\delta)$ yield the
optimal PRCs using Eqs.~\eqref{eq:iPRC_solution} and \eqref{eq:pPRC_solution}.
The maximal parameters $\Delta$ and $\delta$ are calculated numerically.
Figures~\ref{fig:optimal_insolation}(d)--(f) describe the optimal
iPRCs (solid line) and pPRCs (dashed line) for $\alpha=0$, $0.5$,
and $1.0$, respectively. When $\alpha=0$, i.e., the input signal
is additive, $\Psi(\Delta,\delta)$ achieves a maximum for $\Delta=\pi$
and arbitrary $\delta$, yielding sinusoidal PRCs as the optimal solution
(Fig.~\ref{fig:optimal_insolation}(d)). Although the input signal
$p(\phi)$ is not sinusoidal, the optimal PRCs obtained using the
variational method become sinusoidal. In other words, considering
optimality, resonator-type oscillators have an advantage over integrator-type
oscillators. For $\alpha>0$, the input signal $p(\phi)$ depends
on the concentration of the $k$-th molecular species. From Fig.~\ref{fig:optimal_insolation}(b),
the optimal parameters for $\alpha=0.5$ are $(\Delta,\delta)=(2.31,1.99)$
and $(3.98,4.30)$, which are different from $\Delta=\pi$ (these
two sets yield symmetric PRCs with respect to the horizontal axis).
Figure~\ref{fig:optimal_insolation}(e) shows the optimal iPRCs $U_{j}(\phi)$
and pPRCs $Z(\phi)$ for $\alpha=0.5$. Interestingly, the optimal
iPRCs and pPRCs for $\alpha=0.5$ have a dead zone (region of $1\lesssim\phi\lesssim2$
in Fig.~\ref{fig:optimal_insolation}(e)) in which the input signal
neither advances nor delays the clock. From Eqs.~\eqref{eq:iPRC_solution}--\eqref{eq:pPRC_solution}
and the solar radiation input of Eq.~\eqref{eq:insolation_input},
the optimal PRCs inevitably include a dead zone if the optimal $\Delta$
is not $\pi$. For $\alpha=1.0$, there are four sets of parameters
$(\Delta,\delta)$ that give optimal PRCs: $(2.30,2.72)$, $(2.30,1.26)$,
$(3.98,3.56)$, and $(3.98,5.02)$ (PRCs with these four sets are
symmetric each other with respect to the horizontal axis or $\phi=3\pi/2$).
Consequently, the optimal PRCs shown in Fig.~\ref{fig:optimal_insolation}(f)
have a dead zone as in the case of $\alpha=0.5$.

\subsection{Dead zone length}

\begin{figure}
\begin{centering}
\includegraphics[width=13cm]{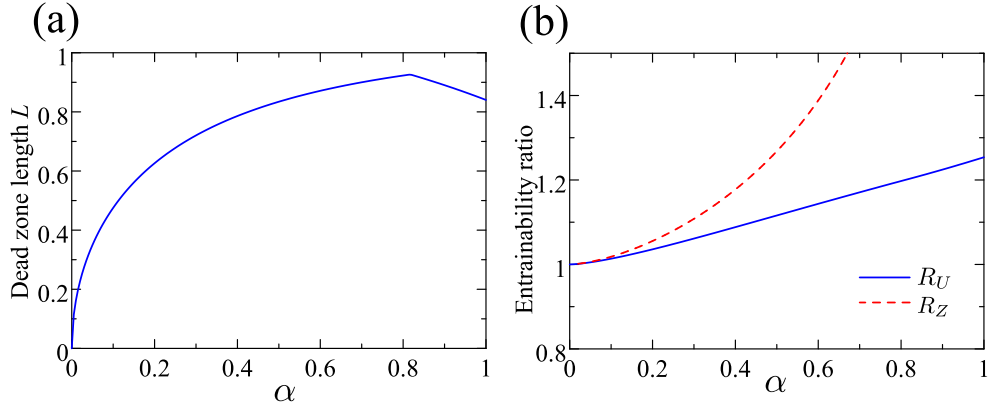} 
\par\end{centering}

\caption{(a) $\alpha$ dependence of the dead-zone length $L$. (b) $\alpha$
dependence of the entrainability ratios $R_{U}$ (solid line) and
$R_{Z}$ (dashed line) (Eq.~\eqref{eq:RU_and_RZ_def}). $R_{U}$
and $R_{Z}$ are the ratios of the entrainability of the optimal PRC
to that of the sinusoidal iPRC (Eq.~\eqref{eq:U_sin_PRC}) and the
pPRC (Eq.~\eqref{eq:Z_sin_PRC}), respectively. \label{fig:deadzone_length}}
\end{figure}

From the results discussed above, the optimal PRCs have a dead zone
when $\alpha>0$. We next studied the length of the dead zone as a
function of $\alpha$ (Fig. \ref{fig:deadzone_length}(a)) and improvements
in the entrainability induced by the dead zone (Fig.~\ref{fig:deadzone_length}(b))
for the solar radiation input. Because the dead zone, which is a null
interval in PRCs, emerges when the optimal parameter is $\Delta\ne\pi$,
we can naturally define its length as 
\begin{equation}
L=|\Delta-\pi|,\label{eq:DZ_length_def}
\end{equation}
where $\Delta$ is the maximum value of $\Psi(\Delta,\delta)$. As
seen in Fig.~\ref{fig:deadzone_length}(a), a dead zone clearly exists
when $\alpha>0$, and the length increases with increasing $\alpha$
for $\alpha<0.8$. Even for $\alpha=0.1$, when the oscillation amplitude
of $x_{\mathrm{LC},k}(\phi)$ (the concentration of a molecular species
modulated by the light-sensitive parameter $\rho$. cf. Fig.~\ref{fig:radiation_fig}(d))
is very small, we observe a dead zone with a length of $L=0.475$,
which corresponds to about 3 h within 24 h, indicating the universality
of having a dead zone in order to attain optimality. The improvement
in the entrainability that is induced by a dead zone is calculated
by comparing the entrainability of the optimal PRCs with that of typical
sinusoidal PRCs. We consider sinusoidal functions for both the iPRC
$U_{j}(\phi)$ and pPRC $Z(\phi)$ by setting 
\begin{eqnarray}
U_{j}(\phi) & \propto & \sin(\phi+c),\label{eq:U_sin_PRC}\\
Z(\phi) & \propto & \sin(\phi+c),\label{eq:Z_sin_PRC}
\end{eqnarray}
where $c$ is the parameter to be optimized so that entrainability
is maximized for each $\alpha$ (see Appendix B for the explicit expressions).
Equations~\eqref{eq:U_sin_PRC} and \eqref{eq:Z_sin_PRC} are scaled
so that they satisfy the constraints on the period variance (Eq.~\eqref{eq:var_T}).
We calculated the ratios 
\begin{equation}
R_{U}=\frac{\mathcal{E}}{\mathcal{E}_{U}},\hspace{1em}R_{Z}=\frac{\mathcal{E}}{\mathcal{E}_{Z}},\label{eq:RU_and_RZ_def}
\end{equation}
where $\mathcal{E}_{U}$ and $\mathcal{E}_{Z}$ represent the entrainabilities
for the cases of the sinusoidal iPRC and pPRC, respectively, calculated
for the solar radiation input. For the sinusoidal iPRC of Eq.~\eqref{eq:U_sin_PRC},
the entrainability is calculated with pPRC via Eq.~\eqref{eq:PRC_Z}.
$R_{U}$ and $R_{Z}$ quantify the improvement rate of the optimal
PRCs over the sinusoidal iPRC ($R_{U}$) and pPRC ($R_{Z}$). In Fig.~\ref{fig:deadzone_length}(b),
the dashed and dot-dashed lines show $R_{U}$ and $R_{Z}$, respectively,
as a function of $\alpha$. Both ratios monotonically increase as
$\alpha$ increases, which shows that the optimal PRC with a dead
zone exhibits better entrainability when the oscillation of $x_{\mathrm{LC},k}(\phi)$
has a larger amplitude. When the concentration of $x_{\mathrm{LC},k}(\phi)$
is low, the effects of the input signal on the circadian oscillators
are smaller. This is because pPRC $Z(\phi)$, which quantifies the
extent of the phase shift due to the stimulation of the parameter,
depends on the concentration $x_{\mathrm{LC},k}(\phi)$ (see Eq.~\eqref{eq:PRC_Z}).
However, even within the range $\phi$ where $x_{\mathrm{LC},k}(\phi)$
has smaller values, the iPRC $U_{j}(\phi)$ contributes to an increase
in the variance of the period, regardless of the concentration. From
this, we see that having an iPRC with a smaller magnitude when the
concentration of $x_{\mathrm{LC},k}(\phi)$ is smaller results in
a smaller variance, which results in a larger entrainability for a
constant variance of the period. Although this qualitatively explains
the benefit of a dead zone, for some input values, the optimal PRCs
may not contain a dead zone for any value of $\alpha$. This will
be shown in the following.

\subsection{Optimal PRC of sinusoidal input}

\begin{figure*}
\begin{centering}
\includegraphics[width=16cm]{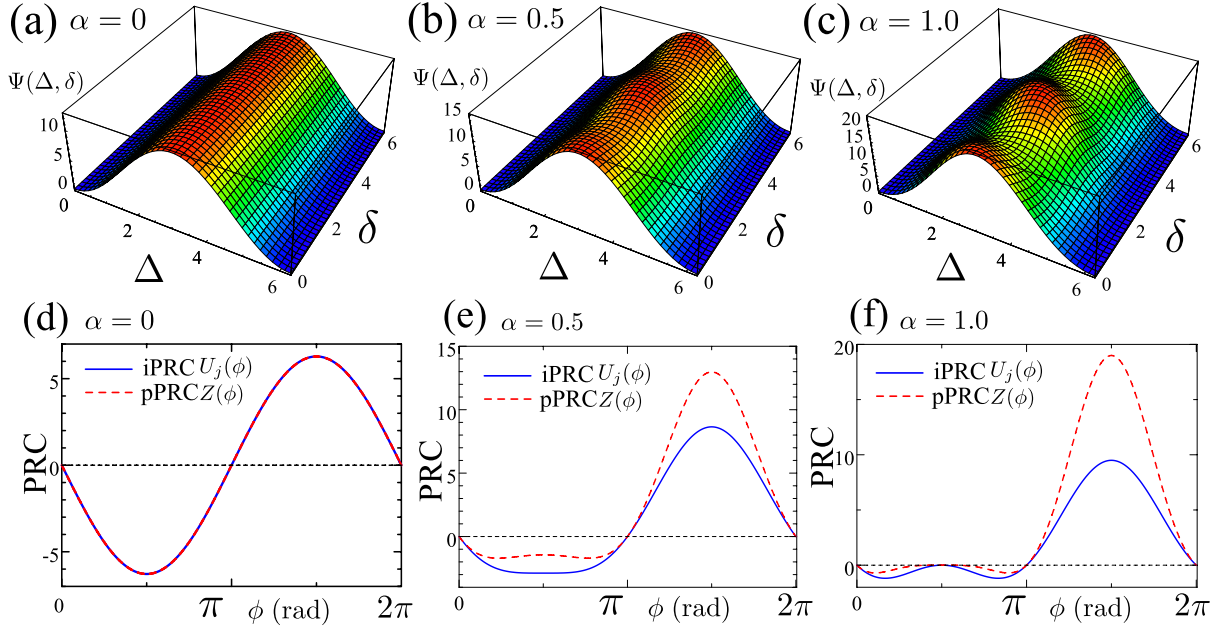} 
\par\end{centering}

\caption{(a)--(c) Landscape of $\Psi(\Delta,\delta)$ as functions of $\Delta$
and $\delta$ with sinusoidal input for (a) $\alpha=0$, (b) $\alpha=0.5$,
and (c) $\alpha=1.0$, where the maximum points are parameters for
the optimal PRC. (d)--(f) Optimal PRCs with sinusoidal input (Eq.~\eqref{eq:sin_input}):
(d) $\alpha=0$, (e) $\alpha=0.5$ and (f) $\alpha=1.0$, where the
solid and dashed lines denote iPRCs $U_{j}(\phi)$ and pPRCs $Z(\phi)$.
The maximal parameter $(\Delta,\delta)$ is $(\pi,0)$ in all cases.
In (d), a parallel shift of the PRC is also optimal ($\delta$ can
be an arbitrary value). PRCs that are symmetric with respect to the
horizontal axis are also optimal. The pPRCs correspond to experimentally
observed PRCs. \label{fig:optimal_sinusoidal}}
\end{figure*}

Since the optimal PRCs depend on input signals (Eqs.~\eqref{eq:iPRC_solution}
and \eqref{eq:pPRC_solution}), we next consider a typical periodic
input signal, a sinusoidal function (Eq.~\eqref{eq:sin_input}).
In this case, $\Psi(\Delta,\delta)$ is calculated in a closed form
(an explicit expression of $\Psi(\Delta,\delta)$ is given in Appendix
A), which is plotted as functions of $\Delta$ and $\delta$ in Fig.~\ref{fig:optimal_sinusoidal}(a)--(c)
for three cases: (a) $\alpha=0$, (b) $\alpha=0.5$ and (c) $\alpha=1.0$.
As can been seen from Fig.~\ref{fig:optimal_sinusoidal}(a)--(c),$\Psi(\Delta,\delta)$
yields the maximal value for $(\Delta,\delta)=(\pi,n\pi)$ for $0<\alpha\le1$,
where $n$ is an integer and when $\alpha=0$, $\delta$ can take
any value. Figures~\ref{fig:optimal_sinusoidal}(d)--(f) express
the optimal iPRCs $U_{j}(\phi)$ and pPRCs $Z(\phi)$ for the sinusoidal
input. For $\alpha=0$, the optimal PRC is sinusoidal (Fig.~\ref{fig:optimal_sinusoidal}(d))
and for $\alpha=0.5$, the optimal PRC is still close to a sinusoidal
function (Fig.~\ref{fig:optimal_sinusoidal}(e)). When increasing
$\alpha$ to $\alpha=1.0$, the PRC diverges from the sinusoidal function
and exhibits almost positive values (Fig.~\ref{fig:optimal_sinusoidal}(f)).
We see that the optimal PRCs due to Eqs.~\eqref{eq:iPRC_solution}
and \eqref{eq:pPRC_solution} do not exhibit a dead zone for any $\alpha$
values (Figs.~\ref{fig:optimal_sinusoidal}(d)\textendash{}(f)) when
the input signal is a simple sinusoidal function.

\section{Discussion}

The existence of a dead zone optimizes both entrainability and regularity.
It is rather obvious that optimization of regularity alone leads to
a dead zone \cite{Pfeuty:2011:EntrainPRC}, because null response
means no effect by any kind of fluctuations. Our result instead shows
that optimality of both entrainability and regularity, which are in
a trade-off relationship, is uniquely achieved by a dead zone. Our
finding is fairly general since a dead zone always exists in an optimal
PRC unless $\alpha=0$ (additive stimulation). Along with the fact
that $T$, $\sigma_{T}$, and $q$ affect only the scaling of the
optimal PRCs, when the input signal affects the dynamics multiplicatively
(i.e., $\alpha>0$), the existence of a dead zone always provides
a synchronization advantage. This is supported by many experimental
studies of various species, that report the existence of a dead zone
in the PRC \cite{Refinetti:2005:CircBook} (cf.~Figs.~\ref{fig:type_1_2_PRCs}(c)--(d)).
Our general result suggests that circadian oscillators have fully
adapted to solar radiation to improve synchronization. Indeed, many
experimental findings imply that circadian oscillators have adapted
to actual solar radiation \cite{Fleissner:2002:NaturalLight}: for
various animals, light-dark (LD) cycles that include a twilight period
result in better entrainability than do abrupt LD cycles (on-off protocols)
\cite{Fleissner:2002:NaturalLight}. In this regard, another interesting
problem is optimal entrainment \cite{Harada:2010:OptimalInput} of
circadian clocks by light stimuli. As two different input signals,
the solar radiation and sinusoidal inputs, yield the same optimal
PRCs for $\alpha=0$, optimal inputs and optimal PRCs do not have
one-to-one correspondence. Thus the optimal inputs are not trivial
and this problem should be pursued in our future studies. 

The solar radiation input plays an essential role, since it yields
a dead zone in the optimal PRC while a sinusoidal signal does not
(see Fig.~\ref{fig:optimal_sinusoidal}). In other words, oscillators
that are entrained by stimuli other than solar radiation may not exhibit
a dead zone in their PRCs. This is indeed found in mammals. Mammals
possess a master clock in their suprachiasmatic nucleus (SCN), which
receives light stimuli via retinal photoreceptors, and peripheral
clocks in body cells \cite{Dibner:2010:MamCircRev}. The peripheral
oscillators are entrained by several stimuli such as feeding and signals
from the SCN through chemical pathways (e.g., hormones) \cite{Dickmeis:2009:GluCircadian,Dibner:2010:MamCircRev}.
By injection experiments of the hormone, Balsalobre \emph{et al.}
\cite{Balsalobre:2000:PeripheralCirc} reported that the PRCs of the
peripheral oscillators in the liver did not have a dead zone.

Our result also agrees with other experimental observations. Our theory
implies that a dead zone should be located where the concentration
$x_{\mathrm{LC},k}(\phi)$ is low ($0\le\phi\le\pi$ in Fig.~\ref{fig:radiation_fig}(d)),
and that to achieve optimality, the concentration of $x_{\mathrm{LC},k}(\phi)$
should be maximal in the region where the PRCs exhibit a large phase
shift. In \emph{Drosophila}, the \textit{timeless} (\emph{tim}) gene
is regarded as the molecular implementation of $x_{\mathrm{LC},k}(\phi)$.
It is experimentally known that light enhances the degradation of
the gene product (the TIM protein) \cite{Hardin:2005:DrosCircReview,Xie:2007:DrosophilaClock},
and the TIM protein peaks during the late evening. Figure~\ref{fig:type_1_2_PRCs}(c)
shows observations of the PRC of \emph{Drosophila} against light pulses
as a function time (hour) from Ref.~\cite{Hall:1987:BiolRhythm};
circles describe the experimental data and the solid line expresses
a trigonometric curve fitting (4th order), respectively. Because the
center of the part of the PRC that can be phase shifted approximately
corresponds to the peak of the concentration, as denoted above, when
estimated from the PRC alone, the concentration peak of the TIM protein
should occur at about 18 h. This time is also close to the experimental
evidence (i.e. late evening). Therefore, our theory can be used to
hypothesize further molecular behavior affected by light stimuli.

In summary, we have constructed a model that regards circadian oscillators
as a global optimization of entrainability and regularity. We have
shown that our model is consistent with much experimental evidence
as mentioned above. The extension and improvement of our method are
possible and they are left as an area of future study.

\appendix

\section{Explicit expression of $\Psi(\Delta,\delta)$}

\subsection{Solar radiation input case}

For the solar radiation input case (Eq.~\eqref{eq:insolation_input}),
$\Psi(\Delta,\delta)$ is given by 
\begin{equation}
\Psi(\Delta,\delta)=\begin{cases}
\Psi_{a}(\Delta,\delta) & 0\le\Delta<\pi,\\
\Psi_{b}(\Delta,\delta) & \pi\le\Delta<2\pi,
\end{cases}\label{eq:Psi_mult_ramp_sin}
\end{equation}
with 
\begin{eqnarray*}
 &  & \Psi_{a}(\Delta,\delta)=\\
 &  & \frac{1}{48q}\left[-3\alpha^{2}\sin(3\Delta+2\delta)+32\alpha\cos(2\Delta+\delta)+12\alpha^{2}\Delta\cos(2\delta+\Delta)\right.\\
 &  & -6\alpha^{2}\sin(2\delta+\Delta)+12\alpha^{2}\pi\cos(\Delta+\delta)^{2}-128\alpha\cos(\Delta+\delta)\\
 &  & +\left\{ -24\alpha^{2}\pi\cos(\delta)^{2}+\left(128\alpha+18\alpha^{2}\sin\delta\right)\cos\delta+(-12\pi+24\Delta)\alpha^{2}-48\pi+48\Delta\right\} \cos\Delta\\
 &  & +12\left(\frac{1}{2}\sin\Delta+\pi\right)\alpha^{2}\cos(\delta)^{2}+\left(24\alpha^{2}\pi\sin\Delta\sin\delta-32\alpha\right)\cos\delta\\
 &  & \left.+\left(-64\alpha\sin\delta-48-27\alpha^{2}\right)\sin\Delta+48\pi+12\alpha^{2}\pi\right],
\end{eqnarray*}
where $\Psi_{b}(\Delta,\delta)=\Psi_{a}(-\Delta+2\pi,-\delta+2\pi)$.
We showed Eq.~\eqref{eq:Psi_mult_ramp_sin} as functions of $\Delta$
and $\delta$ in Fig.~\ref{fig:optimal_insolation}(d)--(f).

\subsection{Sinusoidal input case}

For the sinusoidal input case (Eq.~\eqref{eq:sin_input}), $\Psi(\Delta,\delta)$
is given by 
\begin{equation}
\Psi(\Delta,\delta)=\frac{\pi}{2q}\left(1-\cos\Delta\right)\left[-\alpha^{2}\cos(2\delta+\Delta)+2\alpha^{2}+4\right].\label{eq:Psi_mult_sin}
\end{equation}
We plotted Eq.~\eqref{eq:Psi_mult_sin} as functions of $\Delta$
and $\delta$ in Fig.~\ref{fig:optimal_sinusoidal}(d)--(f).

\section{Explicit expression of sinusoidal PRCs}

\subsection{Sinusoidal iPRC}

An explicit expression sinusoidal iPRC (Eq.~\eqref{eq:U_sin_PRC})
is

\begin{equation}
U_{j}(\phi)=\sqrt{\frac{4\pi^{2}\sigma_{T}^{2}}{qT^{3}}}\sin\left(\phi+c\right),\label{eq:U_sin_PRC_explicit}
\end{equation}
which yields the period variance of $\mathcal{V}_{T}=\sigma_{T}^{2}$.
Then the corresponding pPRC is given by 
\begin{equation}
Z(\phi)=\sqrt{\frac{4\pi^{2}\sigma_{T}^{2}}{qT^{3}}}\sin\left(\phi+c\right)\left(1-\alpha\sin\phi\right),\label{eq:sin_PRC_Z_explicit}
\end{equation}
where we used Eq.~\eqref{eq:PRC_Z}.

\subsection{Sinusoidal pPRC}

For the pPRC $Z(\phi)$ to be a sinusoidal function, the iPRC $U_{j}(\phi)$
must be 
\begin{equation}
U_{j}(\phi)\propto\left(\frac{\partial F_{j}(\phi;\rho)}{\partial\rho}\right)^{-1}\sin(\phi+c),\label{eq:U_pPRC_explicit}
\end{equation}
where we used Eq.~\eqref{eq:PRC_Z}. An explicit expression of Eq.~\eqref{eq:U_pPRC_explicit}
is 
\begin{equation}
U_{j}(\phi)=\frac{1}{\mathcal{N}(c)}\frac{\sin(\phi+c)}{1-\alpha\sin\phi},\label{eq:U_pPRC_def}
\end{equation}
where $\mathcal{N}(c)$ is a normalizing term 
\[
\mathcal{N}(c)=\sqrt{\frac{qT^{3}}{4\pi^{2}\sigma_{T}^{2}}\frac{\alpha^{2}-\left\{ \left(2\sqrt{1-\alpha^{2}}-3\right)\alpha^{2}-2\sqrt{1-\alpha^{2}}+2\right\} \cos(2c)}{\alpha^{2}(1-\alpha^{2})^{3/2}}}.
\]
Equation~\eqref{eq:U_pPRC_def} is normalized so that the period
variance becomes $\mathcal{V}_{T}=\sigma_{T}^{2}$. Using Eq.~\eqref{eq:PRC_Z},
the corresponding pPRC is a sinusoidal function:
\begin{equation}
Z(\phi)=\frac{1}{\mathcal{N}(c)}\sin(\phi+c),\label{eq:Z_2}
\end{equation}
which is an explicit expression of the sinusoidal pPRC (Eq.~\eqref{eq:Z_sin_PRC}).

\section*{Acknowledgment}

This work was supported by the Global COE program ``Deciphering Biosphere
from Genome Big Bang'' from MEXT, Japan (YH and MA); Grant-in-Aid
for Young Scientists B (\#25870171) from MEXT, Japan (YH); Grant-in-Aid
for Scientific Research on Innovative Areas ``Biosynthetic machinery''
from MEXT, Japan (MA)

\end{document}